\documentstyle[aps,prl,twocolumn,epsfig,floats]{revtex}

\begin{document}
\draft \wideabs{
\author{M.V. Romalis, W.C. Griffith, and E.N. Fortson}
\address{Department of Physics, University of Washington, Seattle,
Washington 98195}
\title{A new limit on the permanent electric dipole moment of $^{199}$Hg.}
\date{Submitted Nov 21, 2000}
\maketitle

\begin{abstract}
We present the first results of a new search for a permanent
electric dipole moment of the $^{199}$Hg atom using a UV laser.
Our measurements give $ d(^{199}{\rm Hg})=-(1.06\pm 0.49\pm
0.40)\times 10^{-28}\,e\,{\rm cm}$. We interpret the result as an
upper limit $|d(^{199}{\rm Hg})|<2.1\times 10^{-28}\,e\,{\rm cm}$
(95\% C.L.), which sets new constraints on $\bar{\theta}_{{\rm
QCD}}$, chromo-EDMs of the quarks, and CP violation in
Supersymmetric models.
\end{abstract}
 }

In order for an elementary particle, atom, or molecule to have a
permanent electric dipole moment (EDM) time reversal symmetry must
be violated. By the CPT theorem it also implies a violation of CP
symmetry. A finite EDM would give an unambiguous signal of CP
violation beyond the Standard Model (SM), since EDMs caused by CP
violation in the SM are negligible.  Most extensions of the SM,
such as Supersymmetry, naturally produce EDMs that are comparable
to or larger than present experimental limits \cite{Barr}.
Additional sources of CP violation are motivated by  theories of
baryogenesis \cite{Baryo}.

Experimental searches for EDMs can be divided into three
categories: search for the neutron EDM \cite{nEDM}, search for the
electron EDM utilizing paramagnetic atoms or molecules, the most
sensitive of which is done with Tl atoms \cite{TlEDM}, and search
for an EDM of diamagnetic atoms, the most sensitive of which is
done with $^{199}$Hg \cite{Hg}. The limits set by the most
sensitive experiments in each category are comparable, and they
constrain different combinations of CP-violating effects
\cite{Barr}.

Here we present the first results of a new search for a permanent
EDM of the $^{199}$Hg atom. Using a substantially different
experimental technique we reduce the limit on the $^{199}$Hg EDM
by a factor of 4.  To detect the EDM we measure the Zeeman
precession frequency of $^{199}$Hg nuclear spins ($I=1/2$) in
parallel electric and magnetic fields. The measurements are
simultaneously performed in two cells with oppositely directed
electric fields to reduce the frequency noise due to magnetic
field fluctuations.  A difference between the Zeeman frequencies
in the two cells correlated with reversals of the direction of the
electric field $E$ is proportional to the EDM $d$,
\[
\hbar (\omega _{1}-\omega _{2})=4dE.
\]

\begin{figure}[bt]
\centerline{\psfig{file=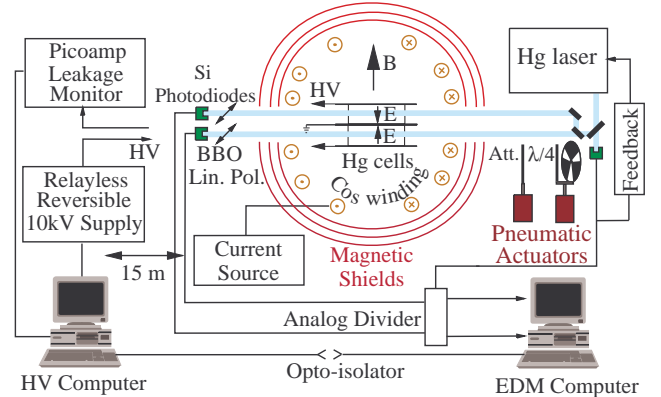,width=3.375in}}
\caption{Schematic of the apparatus used to search for a permanent
EDM of $^{199}$Hg atoms.} \label{app}
\end{figure}

An overall schematic of the apparatus is shown in Figure
\ref{app}. Isotopically enriched $^{199}$Hg vapor (92\%
$^{199}$Hg) was contained in quartz cells with a conductive SnO
coating chemically deposited on the inside surfaces to apply an
electric field. The distance between the electric field plates was
11~mm. A small excess of $^{199}$Hg deposited in the stem of the
cells maintained the number density of $^{199}$Hg atoms close to
the room temperature vapor pressure. The cells also contained
450~torr of N$_{2}$ gas and 50~torr of CO gas. The walls of the
cells were coated with paraffin (C$_{32}$H$_{66}$) to increase the
spin relaxation time. The paraffin was remelted after the cells
were sealed to obtain a thin transparent coating. After such
remelting the $^{199}$Hg spin coherence time was typically about
300-500~sec. However, after a week of continuous UV exposure the
lifetime would drop to below 100~sec. We believe this was due to
damage of the paraffin coating caused by collisions with Hg atoms
in the metastable 6$^{3}P_{0}$ state, to which they are quenched
by N$_{2}$ gas. CO gas is effective in quenching $^{199}$Hg atoms
to the ground state. The spin coherence time could be restored by
remelting the paraffin coating.  The cells were placed in a sealed
vessel made from carbon-filled conductive polyethylene and filled
with SF$_6$ gas. It was located inside a three layer magnetic
shield with a shielding factor of 5$\times 10^{4}$. A magnetic
field of 15~mG was maintained inside the shields by an ultra-low
noise current source \cite{ItaSup}. On a time scale of 100~sec the
field was stable to 25~ppb.

Optical pumping and detection was done using a laser operating at
the 253.7~nm 6$^{1}S_{0}\rightarrow $6$^{3}P_{1}$ transition of
Hg. To generate this wavelength we quadrupled the output of a
semiconductor MOPA (Master Oscillator Power Amplifier) laser
operating at 1015~nm \cite{Stark}. We obtained up to 6~mW of UV
light.  A feedback system adjusted the current of the power
amplifier to keep the light intensity constant. The intensity
noise was $10^{-4}/\sqrt{{\rm Hz}}$ at 10~Hz. The output of the
laser was split into two beams directed perpendicular to the
magnetic and electric fields. For optical pumping the light was
circularly polarized and tuned to the center of the $F=1/2$
hyperfine line of the 6$^{1}S_{0}\rightarrow $6$^{3}P_{1}$
transition. It was chopped at the Larmor frequency of $^{199}$Hg
spins with a duty cycle of 30\%, building-up the polarization in
the rotating frame. To measure the frequency of spin precession
the polarization of the light was switched to linear, the
frequency detuned from resonance by 20~GHz, and the intensity
attenuated to about 7~$\mu $W. Precessing $^{199}$Hg spin
polarization produced an optical rotation of about 60~mrad giving
a 50\% modulation of the intensity transmitted through BBO
Glan-laser polarizers.

A single measurement typically consisted of a 30~sec pump phase
and 100~sec probe phase. During the pump phase the direction of
the electric field was reversed. The high voltage (HV) applied to
each cell was typically alternated between 10~kV and $-10$~kV. A
solid-state relayless HV power supply was used to reduce the
magnetic fields correlated with HV. All HV-related equipment was
located 15~m away from the magnetic shields. We also occasionally
skipped a HV reversal to guard against correlations with periodic
fluctuations. The leakage currents flowing on the walls of the
cells and the vessel were measured using current monitors with
noise less than 0.1~pA. The vessel was designed to provide a
symmetric current path for the charging and leakage currents, so
the magnetic field created by the currents had only a small
projection onto the main magnetic field. The charging currents,
which were on the order of 1~nA, did not produce an observable EDM
signal even when the electric field was reversed during the probe
phase. We also continuously monitored 12 other signals, including
three components of the magnetic field outside of the shields, the
position of the laser beam transmitted through the cell, and
several laser parameters.

A typical run lasted about one day and consisted of several
hundred individual measurements. Each of the spin precession
signals was digitally filtered using a bandpass FFT filter and fit
to an exponentially-decaying sine wave to determine its frequency
and other parameters. The scatter between successive frequency
measurements was due to phase noise and magnetic field noise. We
estimated the contribution from the phase noise by splitting the
signal into short time intervals and fitting them individually. We
verified that the whole detection system was working within 50\%
of fundamental shot-noise limitations. The correlation between the
Zeeman frequency difference and the direction of the electric
field was calculated by analyzing groups of 3 consecutive
measurements and eliminating a linear frequency drift. In most
runs the frequency noise due to magnetic field gradient
fluctuations was comparable to the phase noise, typically
increasing $\chi ^{2}$ for EDM correlations to about 2. The
statistical error was increased by $\sqrt{\chi ^{2}}$ to reflect
actual data scatter in each run.

\begin{figure}[tb]
\centerline{\psfig{file=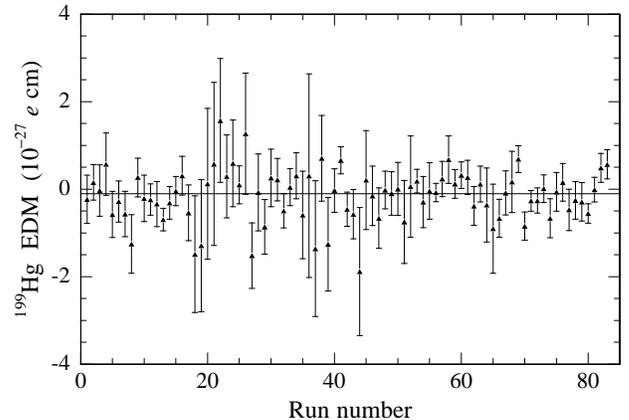,width=3.375in}}
\caption{$^{199}$Hg EDM\ signal as a function of run number. The
solid line shows the average of the data.} \label{sig}
\end{figure}

Frequent reversals and changes were done during the experiment to
monitor for systematic effects. We periodically reversed the data
acquisition channels for the two cells and the direction of the
magnetic field, which should change the sign of the EDM signal. We
also frequently changed the EDM cells and their orientation in the
vessel. In addition, the paraffin in the cells was remelted and
the outside surfaces cleaned each time the cells were changed,
which would likely change the path of the leakage currents. Over
the course of the experiment we used two different vessels and
changed other components of the setup. Figure \ref{sig} shows the
results of all EDM runs. The weighted average of all data gives
$d(^{199}{\rm Hg})=-(1.06\pm 0.49)\times 10^{-28}\,\,e\,{\rm cm}$.
We do not observe any excess data scatter due to changes during
the experiment and the $\chi ^{2}$ per degree of freedom is equal
to 0.95. The statistical error corresponds to a frequency
difference between the two cells of 0.4~nHz, a factor of 5 smaller
than in the previous experiment \cite{Hg}.

\begin{figure}[tb]
\centerline{\psfig{file=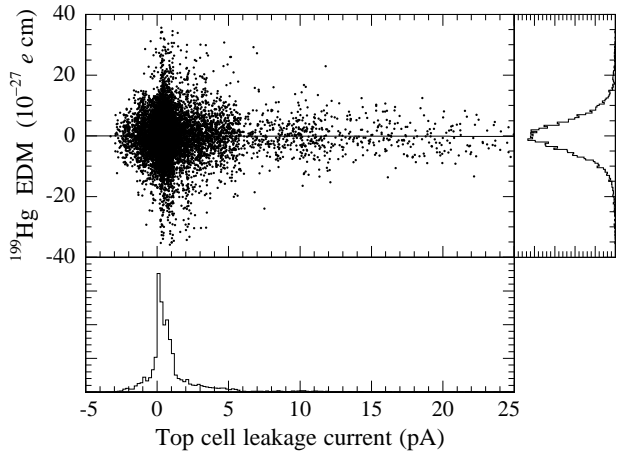,width=3.375in}}
\caption{Correlation between the leakage current and the EDM
signal. Histograms of the leakage current and the EDM data are
also shown. The solid line is a linear fit giving a correlation of
$(-0.4\pm 2.0)\times 10^{-29}e\,{\rm cm/pA}$.} \label{leak}
\end{figure}

\begin{figure}[tb]
\centerline{\psfig{file=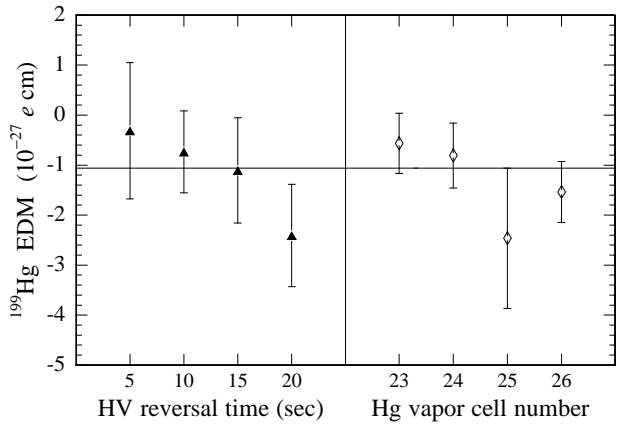,width=3.375in}} \caption{ The
left panel shows the dependence of the EDM signal on the HV
reversal time. The right panel shows the EDM signal obtained with
each of the EDM cells. The solid line is an average of all data.}
\label{dep}
\end{figure}

We looked for systematic effects by changing the operating
parameters of the experiment, exaggerating certain imperfections,
and looking for correlations among different parameters. The
leakage currents are a potentially serious source of systematic
errors because they can produce magnetic fields that are
correlated with the electric field and mimic an EDM signal. It
should be noted that only leakage currents flowing in a helical
path around the cell will contribute to first order. Figure
\ref{leak} shows a scatter plot of the EDM signal vs.\ the leakage
current in one of the cells. No statistically significant
correlation was observed. The average cell leakage currents were
about 0.6~pA. From the error on the correlation slope we can set a
limit on the contribution of the leakage current to the EDM signal
of $0.14\times 10^{-28}\,e\,{\rm cm}$. We estimate the error more
conservatively by calculating the magnetic field created by a
leakage current making one complete loop around the cell. This
rather unlikely path would give an average EDM signal of
$0.25\times 10^{-28}\,\,e\,{\rm cm}$. A total of 4 vapor cells
were used in the experiment in various pairs. The right panel of
Figure \ref{dep} shows that the EDM data taken with each cell are
consistent. Note that if a cell had a fixed helical path for the
leakage current, it would produce the same EDM signal independent
of the orientation of the cell. As can be seen in Figure
\ref{leak} , the leakage currents were sometimes negative. We
believe this effect was due to changes in the mutual capacitance
caused by redistribution of charges on HV insulators. If the HV
was not reversed for a long time, the leakage currents became
positive and approached a steady state value of about 0.1~pA.

We looked for correlations with the electric field of 30 other
variables, such as monitored signals and fitting parameters, and
found no statistically significant correlations. Using random
fluctuations of the variables we determined the cross-correlation
between each of them and the EDM signal $\omega _{1}-\omega _{2}$.
In this way we set upper limits on false EDM signals coming from
cross-correlations. All these limits are 10 to 100 times smaller
than our statistical error. For positive direction of the magnetic
field the average EDM signal was $ d(B+)=-(1.78\pm 0.70)\times
10^{-28}\,\,e\,{\rm cm}$ and for negative direction
$d(B-)=-(0.36\pm 0.69)\times 10^{-28}\,\,e\,{\rm cm}$. The two
results are within 1.4~$\sigma $ of each other. A systematic
effect that does not reverse with the magnetic field would show up
in the difference but cancel in the average of the two results. To
study possible frequency shifts due to magnetization of the
magnetic shields caused by the charging currents, we varied the
high voltage reversal time from 5 to 20~sec. The dependence of the
EDM signal on the HV reversal time, shown in the left panel of
Figure \ref{dep},  is not statistically significant. We did not
resolve any correlations of the individual Larmor frequencies
$\omega _{1}$ and $\omega _{2}$ with the electric field outside of
their error bars, which are a factor of 6 larger than the
statistical error on $\omega _{1}-\omega _{2}$.

We looked for effects proportional to $E^{2}$ in separate runs by
applying the electric field to only one of the two cells and
alternating the HV between 0 and $\pm 10$~kV. The quadratic
frequency shift was less than 2~nHz. We checked that the electric
field in the cells was uniform and reversible with an accuracy of
1.5\% \cite{Stark}, which limits the effect of reversal
imperfections to less than $7\times 10^{-30}e\,{\rm cm.}$ Although
the average velocity of the atoms in the cell is equal to zero,
residual $ v\times E$ effects \cite{TlEDM} can exist if the
surface relaxation on the walls is asymmetric. We looked for these
effects by taking data with the magnetic field intentionally
misaligned by 5$^{\circ }$ from the electric field. No effects
were seen at the level of $1.5\times 10^{-28}e\, {\rm cm,}$ which
can be used to constrain this effect to less than $0.3\times
10^{-28}e\,{\rm cm} $ in a magnetic field aligned within 1$^{\circ
}$ relative to the electric field. Among various frequency shifts
caused by the probe light the most significant is due to the
magnetic dipole and electric quadrupole transitions in an electric
field \cite{MD}. This effect is odd in the $E$ field and can mimic
an EDM signal. It is suppressed to about  $3\times
10^{-30}\,\,e\,{\rm cm}$ because the laser beam is directed
perpendicular to the magnetic field and detuned far from
resonance.

In summary, no statistically significant systematic effects that
mimic an EDM signal were observed, although in several cases our
systematic studies were limited by statistics. We estimate the
total systematic uncertainty to be $0.40\times 10^{-28}\,e\,{\rm
cm}$ by adding in quadrature the limits on systematic effects due
to the leakage currents, the $v\times E$ effect, and other
miscellaneous effects. Thus we obtain $d(^{199}{\rm Hg})=-(1.06
\pm 0.49\pm 0.40)\times 10^{-28}\,e\,{\rm cm}$ and interpret the
result as an upper limit on the $^{199}$Hg EDM $|d(^{199}{\rm
Hg})|<2.10\times 10^{-28}\,e\,{\rm cm}$ (95\% C.L.).

This limit can be used to place new constraints on hadronic and
semi-leptonic CP-violating effects which are summarized in Table
\ref{table}. The EDM of the $^{199}$Hg atom is proportional to the
Schiff moment of the $^{199}$Hg nucleus $S$, which is a measure of
the difference between the distributions of the electric charge
and electric dipole moment in the nucleus. Using a Hartree-Fock
calculation for Hg atomic wavefunctions \cite {Pend} and a simple
nuclear shell model \cite{Flam,EDMbook} the Schiff moment has been
calculated with an uncertainty of about 30-50\%: $ d(^{199}{\rm
Hg})=-3.1\times 10^{21}S$~cm$^{-2}$\cite{EDMbook}. The largest
contribution to the Schiff moment comes from a CP-violating
nucleon-nucleon interaction $\xi G_{F}(\bar{p}p)(\bar{n}i\gamma
_{5}n)/\sqrt{2}$. It was calculated in \cite{Flam} using
Woods-Saxon potentials and neglecting many-particle correlations.
The result is $S=-1.8\times 10^{-7}\xi \,e\,$fm$ ^{3}$ with an
uncertainty of about 50\%. Possible enhancements of the Schiff
moment due to collective octupole nuclear excitations have been
considered recently in \cite{Hayes}, although no definite
estimates exist. As shown in \cite{Khatim,Posp}, the CP-odd
nucleon-nucleon interaction is dominated by $ \pi ^{0}$ exchange
and is proportional to the pion-nucleon CP-odd coupling constant
$\bar{g}_{\pi NN}$.

\begin{table}[tbp] \centering%
\begin{tabular}{lr@{$\times$}lr@{$\times$}l@{}rr}
Parameter & \multicolumn{2}{c}{Limit from $^{199}$Hg} &
\multicolumn{3}{c}{ Best other limit} & Th.\ Ref.\\
\hline $\bar{\theta}_{\rm QCD}$ & $1.5$ & $10^{-10}$ & $6$ &
$10^{-10}$ & n  \cite{nEDM} & \cite{Posp,Pospth}
\\
$\tilde{d}_{d}\,({\rm cm)}$ & $7$ & $10^{-27}$ & $1.1$ & $10^{-25}$ &n \cite{nEDM} & \cite{Posp,Pospn} \\
$C_{T}$ & $1$ & $10^{-8}$ & $5$ & $10^{-7}$ & TlF  \cite{TlF}& \cite{EDMbook} \\
$C_{S}$ & $3$ & $10^{-7}$ & $4$ & $10^{-7}$ & Tl  \cite{TlEDM} & \cite{EDMbook} \\
$\varepsilon _{q}^{\rm SUSY}$ & $2$ & $10^{-3}$ & $1$ & $10^{-2}$
&n   \cite{nEDM}& \cite{Barr}
\\
$\varepsilon ^{\rm Higgs}$ & \multicolumn{2}{l}{$0.4/{\rm
tan}\beta $} &
\multicolumn{2}{l}{$0.7/{\rm tan}\beta $} &Tl  \cite{TlEDM} & \cite{Barr} \\
$x^{\rm LR}$ & $1$ & $10^{-3}$ & $1$ & $10^{-2}$ &n  \cite{nEDM}&
\cite{Barr}
\end{tabular}

\caption{Summary of limits (95\%C.L.) set by the  $^{199}$Hg EDM
and other experiments on model-independent and ``naturalness''
parameters.} \label {table}
\end{table}%

A limit on $\bar{g}_{\pi NN}$ can be used to directly constrain
the CP-violating QCD vacuum angle $\bar{\theta}_{{\rm QCD}}$
\cite{Witten}. We obtain $|\bar{\theta}_{{\rm QCD}}|<1.5\times
10^{-10}$, improving the limit set by the neutron EDM
\cite{nEDM,Pospth} by a factor of 4. We can also set a limit on a
linear combination of chromo-EDMs of the quarks\cite{Posp},

\[
e|\tilde{d}_{d}-\tilde{d}_{u}-0.012\tilde{d}_{s}|<7\times
10^{-27}e\,{\rm cm.\,}
\]
This limit can be compared with a constraint on a different
combination of EDMs and chromo-EDMs set by the neutron EDM
experiment\cite{nEDM,Pospn},
\[
|e(\tilde{d}_{d}+0.5\tilde{d}_{u})+1.3d_{d}-0.3d_{u}|<1.1\times
10^{-25}e\, {\rm cm.}
\]
In most extensions of the SM, including Supersymmetry, EDMs and
chromo-EDMs of the quarks have comparable size \cite{Barr}. We
also place new constraints on semileptonic CP-violating parameters
$C_{S}$ and $C_{T}$, which are significant for certain multi-Higgs
models\cite{Barrhiggs}.

In addition to the model-independent constraints discussed above,
one can set limits on specific CP-violating parameters in various
extensions of the SM. For example, in the Minimal Sypersymmetric
SM the limit on the $^{199}$Hg EDM can be used to set tight
constraints on a linear combination of two CP-violationg phases
\cite{Posp}. In Table \ref{table} we only give general limits for
``naturalness'' parameters, as defined in \cite{Barr}, for
Supersymmetric, multi-Higgs, and Left-Right symmetric models. For
example, in Supersymmetry $\varepsilon _{q}^{\rm SUSY}$ would be
close to unity if the masses of sypersymmetric particles were on
the order of 100~GeV and CP-violating phases were large.

In conclusion, we have presented the results of a new search for a
permanent electric dipole moment of $^{199}$Hg atoms, improving
the previous limit by a factor of 4. We have set new limits on
$\bar{\theta}_{\rm QCD}$, quark chromo-EDMs, and CP violation in
various extensions of the Standard Model. We are presently
upgrading the experiment and plan to improve the statistical
sensitivity by at least a factor of 2. We would like to thank
Warren Nagourney for help with the laser quadrupling system, Jim
Jacobs for assistance in fabrication of the EDM cells, and Blayne
Heckel for helpful discussions. This work was supported by NSF
Grant No. PHY-9732513.

\end{document}